\documentclass[11pt,twoside]{article}
\usepackage{asp2006}
\usepackage{psfig}
\usepackage{epsf}
\usepackage{graphicx}
\usepackage{lscape}
\newcommand{\begit}{\begin{itemize}}
\newcommand{\enit}{\end{itemize}}
\newcommand{\beq}{\begin{equation}} 
\newcommand{\eeq}{\end{equation}} 
\newcommand{\beqa}{\begin{eqnarray}} 
\newcommand{\eeqa}{\end{eqnarray}}

\newcommand{\sds}{\dot{\Sigma}_\star}

\newcommand{\begen}{\begin{enumerate}}
\newcommand{\enen}{\end{enumerate}}
\def\lesssim{\mathrel{\hbox{\rlap{\hbox{\lower4pt\hbox{$\sim$}}}\raise2pt\hbox{$<$}}}}
\def\gtrsim{\mathrel{\hbox{\rlap{\hbox{\lower4pt\hbox{$\sim$}}}\raise2pt\hbox{$>$}}}}

\def\edcomment#1{\iffalse\marginpar{\raggedright\sl#1\/}\else\relax\fi}
\reversemarginpar

\begin{document}
\title{Radiation Pressure Supported Starburst Galaxies
\& The Fueling of Active Galactic Nuclei}
\author{Todd A.~Thompson}
\affil{Department of Astronomy and 
Center for Cosmology \& Astro-Particle Physics,
The Ohio State University, Columbus, Ohio 43210; Email: 
thompson@astronomy.ohio-state.edu}

\begin{abstract}
Radiation pressure from the absorption and scattering
of starlight  by dust grains may be a crucial feedback
mechanism in starburst galaxies and the self-gravitating 
parsec-scale disks that accompany the fueling of active galactic 
nuclei. I review the case for radiation pressure in both 
optically-thin and highly optically-thick contexts.  I describe the
conditions for which Eddington-limited star formation 
yields a characteristic flux of $\sim10^{13}$\,L$_\odot$ kpc$^{-2}$,
and I discuss the physical circumstances for which the flux
from radiation pressure supported disks is below or above
this value.  In particular, I describe the young stellar disk 
on $\sim0.1$\,pc scales at the Galactic Center. I argue that its 
bolometric flux at formation, $\sim10^{15}$\,L$_\odot$ kpc$^{-2}$,
and the observed stellar mass and scale height imply that 
the disk may have been  radiation pressure supported during formation.
\end{abstract}

\section{Introduction}

Star formation in galaxies is observed to be globally inefficient,
with only a few percent 
of the available gas supply converted to stars per dynamical 
timescale (e.g., Kennicutt 1998).
On smaller scales within 
individual galaxies star formation is similarly 
slow (e.g., Kennicutt et al.~2007; Bigiel et al.~2008; 
Leroy et al.~2008; Krumholz \& Tan 2007).  
This, together with the fact that galaxies are 
marginally-stable, with Toomre's $Q\sim1$ (e.g., Martin \& Kennicutt 2001), 
suggests that they are self-regulated --- otherwise,
disks should quickly evolve to have  $Q < 1$, and 
fragment on a single dynamical time.

The injection of energy and momentum into the 
interstellar medium (ISM) by stellar processes (``feedback''), 
may be responsible for the inefficiency of 
star formation in galaxies and for their self-regulation. 
The feedback mechanism thought to stave off self-gravity 
is uncertain, but may be due to 
supernovae, stellar winds, expanding HII regions, 
cosmic rays, and radiation pressure of starlight on dust  
(e.g., McKee \& Ostriker 1977; Chevalier \& Fransson 1984; 
Silk 1997; Wada \& Norman 2001; Scoville et al.~2001; Matzner 2002; 
Thompson, Quataert, \& Murray 2005 [TQM]; 
Li \& Nakamura 2006; Socrates et al.~2007). 
Non-stellar processes can also act as a feedback mechanism by 
tapping the galaxy's gravitational binding energy 
(Sellwood \& Balbus 1999).

Both observations and theory suggest that whatever the mechanism, 
feedback must be yet more effective on parsec scales 
around active galactic nuclei (AGN), at surface densities
in excess of those probed by the densest galaxies
that comprise the Schmidt Law.  On scales 
larger than $\sim0.01$\,pc accretion disks feeding bright AGN are
gravitationally unstable and prone to fragmentation and
star formation (Kolykhalov \& Sunyaev 1980;
Shlosman et al.~1989, 1990; Kumar 1999; Wada \& Norman 2002; Goodman 2003).  This problem is 
severe: gas accumulating on $\sim0.01-{\rm few}$\,pc scales
to fuel a supermassive black hole may never get there, but instead 
fragment entirely into stars.  Both rapid inflow due to efficient
angular momentum transport (e.g., bars) and/or a reduced
star formation efficiency have been proposed to help solve this 
``AGN fueling problem'' (Shlosman et al.~1990; Levin 2007; TQM).  

A number of observational results also suggest an intimate connection
between AGN fueling and nuclear star formation. 
For example, Storchi-Bergmann et al.~(2005) and
Davies et al.~(2006), (2007) find direct evidence
for star formation in the central tens of parsecs of local AGN.  
Likewise, the young stellar disk on $\sim0.1$\,pc scales in the 
Galactic Center (e.g., Paumard et al.~2006; Lu et al.~2008) may be 
a vestige of accretion,  providing further evidence for a close link between
BH fueling and star formation (Levin \& Beloborodov 2003;
Milosavljevic \& Loeb 2004; TQM; Nayakshin et al.~2007).  There is 
further evidence that the fueling of bright high-$z$ quasars
is accompanied by intense star formation (e.g., Lutz et al.~2007, 2008;
Younger et al.~2008).

In this contribution, I review the arguments for the importance of 
radiation pressure of starlight on dust grains as an important feedback
mechanism in intense starburst galaxies and in the  
self-gravitating disks that likely attend AGN fueling.  
In \S\ref{section:radiation}, I
discuss some of the physics of this particular feedback mechanism.
In \S\ref{section:examples}, I discuss some examples of
starburst systems for which the Eddington limit may be relevant,
including local ULIRGs and the young stars at the Galactic Center.
Section \ref{section:summary} provides a brief conclusion.
Throughout, I borrow from the more extensive discussion 
presented in TQM. The interested reader is referred 
there for more details.

\section{Radiation Pressure Feedback}
\label{section:radiation}

The light from massive stars is efficiently absorbed and
scattered by dust grains, providing a net force in the direction
normal to a star-forming disk. TQM showed that this might be the 
dominant feedback mechanism in dense optically-thick starbursts
and AGN disks (see also Scoville et al.~2001; Scoville 2003; Levin 2007).    
A basic requirement for
radiation pressure to be important in governing the dynamics 
is that the radiative flux approach the  Eddington flux
\beq
F_{\rm Edd}\approx\frac{4\pi G\Sigma_{\rm tot}c}{\langle\kappa_F\rangle}
\approx6\times10^{13}\,\,\frac{{\rm L_\odot \,\,kpc^{-2}}}{\rm g^2\,\,\,cm^{-4}}
\left(\frac{\Sigma_{\rm tot}}{\langle\kappa_F\rangle}\right),
\label{edd}
\eeq
where $\Sigma_{\rm tot}$ is the total disk surface density and
$\langle\kappa_F\rangle$ is the column-averaged flux-mean
dust opacity (e.g., Elitzur \& Ivezic 2001).
Note that we expect the gas and 
dust to behave as a single collisionally-coupled fluid on the scales
and at the densities of interest (e.g., Laor \& Draine 1993; 
Murray et al.~2005 [MQT]).
Because the radiative flux from galaxies is proportional to the
star formation rate per unit area ($F_\star=\epsilon\sds c^2$),\footnote{$\epsilon\approx10^{-3}$
is an IMF-dependent constant; see Kennicutt (1998).}
equation (\ref{edd}) is an interesting and testable form for the Schmidt Law:
\begin{equation}
\dot{\Sigma}_{\rm \star,\,Edd}\approx\frac{4\pi G}{\epsilon c}
\left(\frac{\Sigma_{\rm tot}}{\langle\kappa_F\rangle}\right)
\approx4000\,\,\frac{{\rm M_\odot\,\,yr^{-1}\,\,kpc^{-2}}}{{\rm g^2\,\,\,cm^{-4}}}
\left(\frac{\Sigma_{\rm tot}}{\langle\kappa_F\rangle}\right).
\label{sds}
\end{equation}
Although $\Sigma_{\rm tot}$ is relatively easy to measure or
estimate in normal star-forming galaxies and starbursts, the coupling
between the radiation field and the dust, as expressed in equation
(\ref{edd}) with $\langle\kappa_F\rangle$ is more difficult, since
the medium is highly turbulent, inhomogeneous, and clumpy.
Three simple limits help to illustrate the range of the
effective flux-mean opacity in galactic contexts.

\subsection{Optically-Thin to UV: $\Sigma_g\lesssim5$\,M$_\odot$ pc$^{-2}$ $\approx10^{-3}$\,g cm$^{-2}$}

When the average medium is optically-thin to the UV radiation from massive 
stars one finds that for young stellar populations ($\lesssim5$\,Myr), 
standard grain size distributions, and Galactic dust-to-gas ratio
that $\langle\kappa_F\rangle\sim10^3$\,\,cm$^2$ g$^{-1}$.  This 
limit is only applicable for galaxies with gas surface densities
of $\Sigma_g\lesssim5$\,M$_\odot$ pc$^{-2}$, again assuming Galactic
dust-to-gas ratio. For typical numbers, 
\begin{equation}
F_{\rm \star,\,Edd}
\approx6\times10^{7}\,\,{\rm L_\odot\,\,kpc^{-2}}
\left(\frac{\Sigma_{\rm tot}}{5\,\,{\rm M_\odot\,\,pc^{-2}}}\right)
\left(\frac{10^3\,\,{\rm cm^2 \,\,g^{-1}}}{\langle\kappa_F\rangle}\right).
\end{equation}

\subsection{Optically-Thick to UV, but Optically-Thin to Re-Radiated FIR: 
$5\,{\rm M_\odot\,\,pc^{-2}}\lesssim\Sigma_g\lesssim5000$\,M$_\odot$ pc$^{-2}$}

In normal star-forming galaxies and some starbursts, 
the average gas surface density $\Sigma_g$ is high enough that the medium is 
optically-thick to the UV emission from massive stars,
but not sufficiently large that 
the re-radiated FIR emission from dust grains is optically-thick.  

In this ``single-scattering'' limit (see MQT, TQM), UV photons are
absorbed once, and then escape the system.  For a homogeneous slab
of gas and dust and with an incident UV radiation field, it is easy
to show that $\langle\kappa_F\rangle\sim2/\Sigma_g$ (e.g., TQM).  
Although the applicability of the homogeneous slab is highly 
uncertain in turbulent star-forming environments, the single-scattering
approximation is useful at the order-of-magnitude level for gauging
the dynamical importance of the radiation field.  In this case, 
\beq
F_{\star,\,\rm Edd}\approx
2\pi G\Sigma_{\rm tot}\Sigma_g c 
\sim10^{9}\,\,{\rm L_\odot \,\,kpc^{-2}}
\left(\frac{\Sigma_{\rm tot}}{100\,\,{\rm M_\odot\,\,pc^{-2}}}\right)
\left(\frac{\Sigma_g}{10\,\,{\rm M_\odot\,\,pc^{-2}}}\right).
\label{thin}
\eeq

\subsection{The Optically-Thick Limit: $\Sigma_g\gtrsim5000$\,M$_\odot$ pc$^{-2}$}

When the galaxy is optically-thick
to the re-radiated FIR emission from dust grains,  $\langle\kappa_F\rangle$
can be approximated by the Rosseland-mean opacity, $\kappa_R(T)$,
where $T$ is the midplane temperature of the star-forming disk.
Although the temperature dependence of the opacity is fairly complicated,
to rough approximation when $T\lesssim200$\,K, $\kappa_R\approx2\times10^{-4}T^2\equiv\kappa_0T^2$, 
and for $200\,{\rm K}\lesssim T\lesssim T_{\rm sub}$, $\kappa_R(T)\approx
{\rm constant}\approx5-10$\,cm$^2$ g$^{-1}$ (e.g., Bell \& Lin 1994; Semenov et al.~2003).
For temperatures above the sublimation temperature of dust 
$T_{\rm sub}\approx 1500$\,K, the Rosseland mean opacity 
decreases markedly. \\

{\noindent \bf The Eddington Limit when $T\lesssim200$\,K:}
In this limit, the midplane temperature is connected 
to the effective temperature and the radiated flux by 
$T^4\sim\tau T_{\rm eff}^4\sim(\kappa_0T^2\Sigma_g/2)(F/\sigma_{\rm SB})$.
Additionally, if the radiation pressure
balances gravity, $p_r=aT^4/3\sim\pi G\Sigma_g\Sigma_{\rm tot}$.
These two equations imply that
\beq
F_{\star,\,\rm Edd}\sim(3\pi c G\sigma_{\rm SB}/\kappa^2_0)^{1/2}
\left(\Sigma_{\rm tot}/\Sigma_g\right)^{1/2} 
\sim10^{13}\,\,{\rm L_\odot\,\,kpc^{-2}} \,
\left(\Sigma_{\rm tot}/\Sigma_g\right)^{1/2},
\label{flux}
\eeq
where $\kappa_0=2\times10^{-4}$\,cm$^{2}$ g$^{-1}$ K$^{-2}$ has been assumed.
For typical values of $\epsilon$ (eq.~\ref{sds}), equation (\ref{flux}) predicts a  
star formation rate surface density of 
approximately $\dot{\Sigma}_\star\sim10^3$\,M$_\odot$ yr$^{-1}$ kpc$^{-2}$ required
to support the disk with radiation pressure on dust.
Note that because we expect starbursts that meet the criteria used to derive
equation (\ref{flux}) to be largely gas-dominated, $\Sigma_{\rm tot}\sim\Sigma_g$
is a reasonable first approximation.  Hence, 
starburst galaxies with $\Sigma_g\gtrsim5000$\,M$_\odot$ pc$^{-2}$ 
should attain a characteristic flux of $\sim10^{13}$\,L$_\odot$ kpc$^{-2}$. 

The primary assumptions made in deriving
equation (\ref{flux}) are that the medium is optically-thick and that the 
midplane temperature is less than $\sim200$\,K.  For example, a massive 
starburst with a gas reservoir of $\sim10^{10}$\,M$_\odot$ distributed on 
scales $\lesssim1$\,kpc should satisfy both criteria over approximately 
1 decade in radius, down to scales of $\lesssim100$\,pc, depending on the 
radial distribution of the stellar and gas mass.  If the surface density
increases at smaller radii, the midplane temperature will as well and 
the assumption that $T\lesssim200$\,K will be violated.  Similarly, on
somewhat larger scales, the medium is optically-thin to the 
re-radiated FIR emission and equation (\ref{thin}) becomes applicable.\\

{\noindent \bf The Eddington Limit when $T\gtrsim200$\,K:}
If the equilibrium radiative
flux required for hydrostatic equilibrium implies $200{\rm \,K}\lesssim T\lesssim T_{\rm sub}$,
then $F_{\rm Edd}$ exceeds the value given in equation (\ref{flux}).
In this case, we can approximate $\kappa_R(T\gtrsim200\,\,{\rm K})\sim5-10$\,cm$^2$ g$^{-1}$
and
\begin{equation}
F_{\star,\,\rm Edd}\sim 10^{15}\,\,{\rm L_\odot\,\,kpc^{-2}}
\,\left(\frac{10\,{\rm cm^2\,\,g^{-1}}}{\kappa_R}\right)
\left(\frac{\Sigma_{\rm tot}}{10^6\,\,{\rm M_\odot\,\,pc^{-2}}}\right),
\label{high}
\end{equation}
where I have scaled the total surface density for parameters 
characteristic of $\sim0.1-1.0$\,pc self-gravitating disks 
around AGN (see \S\ref{section:gc}).

\subsection{Stability}

Equation (\ref{flux}) implies that if an optically-thick starburst
with $T\lesssim200$\,K reaches a characteristic flux of $\sim10^{13}$\,L$_\odot$ kpc$^{-2}$,
then it can be maintained in hydrostatic equilibrium in the sense that 
radiation pressure balances the self-gravity of the disk.  However, 
because the diffusion timescale for radiation is very short compared 
to the dynamical timescale, the disk is unstable.  In
other words, radiation pressure cannot stave off the local Jeans 
instability on the scale of the disk gas scale-height because
diffusion is rapid (see Thompson 2008).   This effect should keep the
disk highly turbulent, with the magnitude of the resulting turbulent 
energy density ($\rho \,\delta v^2$) set by the magnitude of the 
radiation energy density.  In this way, hydrostatic equilibrium 
would be maintained statistically.  
As discussed in Thompson (2008), turbulence may also generate the
$\sim{\rm few}$\,mG magnetic field strengths needed in the densest
starbursts to explain the fact that these systems lie on the FIR-radio 
correlation (Thompson et al.~2006).

\section{Some Examples}
\label{section:examples}

\subsection{Local \& High-$z$ Starbursts}

\begin{figure}
\centerline{\includegraphics[width=6.2cm]{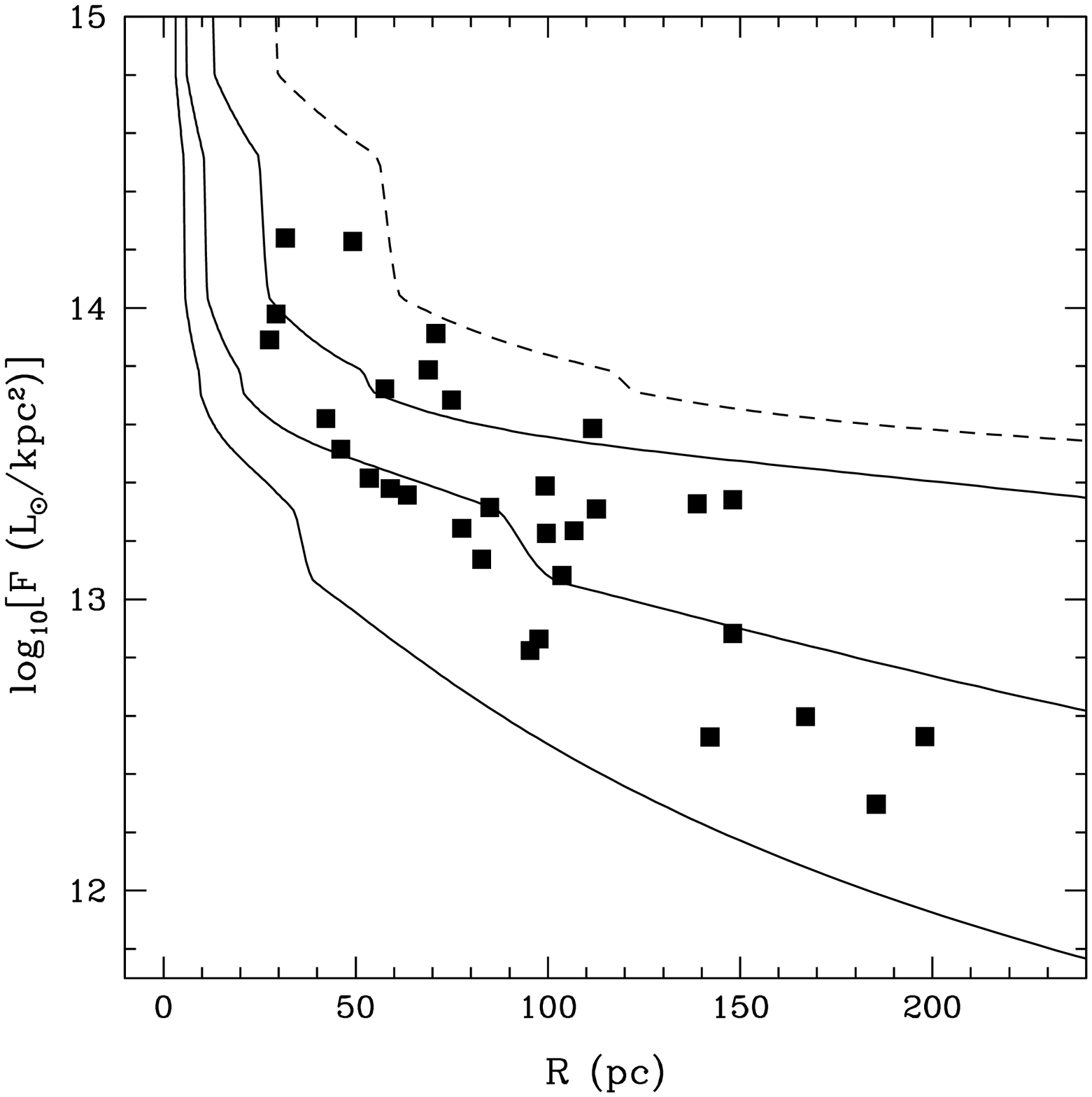}
\includegraphics[width=6.2cm]{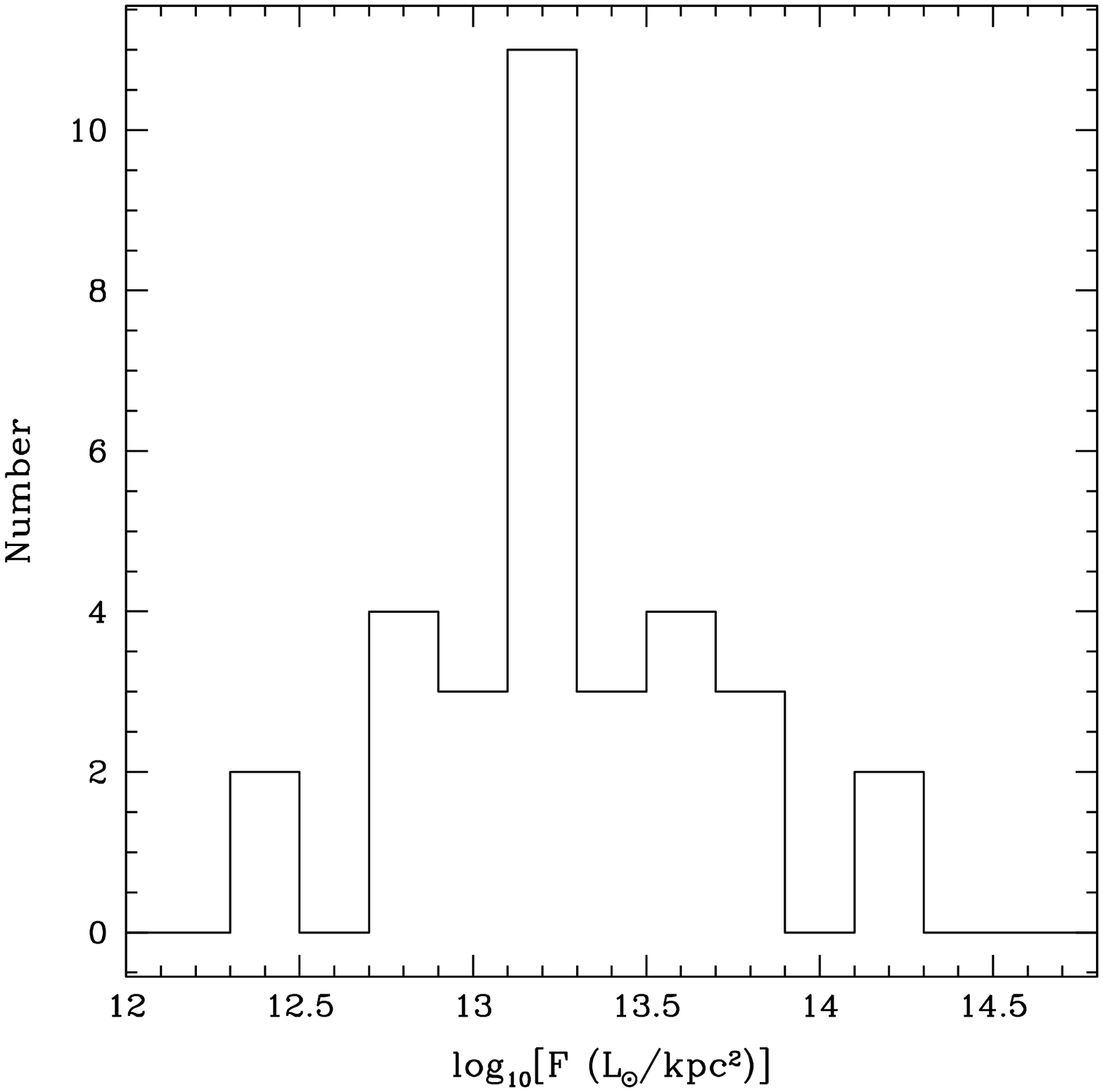}}
\caption{{\it Left:}
Flux as a function of radius, as computed by dividing the 
FIR luminosity by the radio continuum area, for the local ULIRGs 
in Condon et al.~(1991; filled squares) compared with a set of theoretical 
models from TQM using equations (\ref{flux}) and (\ref{high})
(solid \& dashed lines).  The solid lines have constant gas
fraction of 0.1, 0.3, and 1.0 from lowest to highest.  The dashed 
line shows a model with a deeper stellar potential.
{\it Right:} Histogram showing number of ULIRGs as a function of flux.
The peak at $\sim10^{13}$\,L$_\odot$ 
kpc$^{-2}$ is consistent with the prediction of 
Eddington-limited star formation in radiation 
pressure supported self-gravitating disks.  Plots 
from TQM.}
\label{condon}
\end{figure}

The flux {\it predicted} by equations (\ref{flux})
and (\ref{high}) can be compared with observations of local 
and high-$z$ ultra-luminous infrared galaxies.  Figure \ref{condon}
({\it left}) shows a comparison between the observed fluxes of 
local ULIRGs and a set of models with increasing constant gas
fraction from TQM.  The right panel displays the same data as 
a histogram, and indicates that ULIRGs may in fact exhibit a
characteristic flux of $\sim10^{13}$\,L$_\odot$ kpc$^{-2}$.
It is also interesting in this context to note that 
F\"orster-Schreiber et al.~(2003) find that the 
$\sim10$\,Myr starburst required to explain the 
observed stellar population in the central starburst
of M82 also attained a star formation rate surface 
density of $\dot{\Sigma}_\star\sim10^3$\,M$_\odot$ yr$^{-1}$ 
kpc$^{-2}$ (see their Fig.~13).

Finally, observations of submillimeter galaxies at high-$z$ are also in 
fair agreement with the predictions of TQM;
Younger et al.~(2008), Walter et al.~(2009),
and Riechers et al.~(2009)
argue that the luminosities of the systems they observe indicate
that radiation pressure on dust may be dynamically important.

\subsection{The Galactic Center \& The Fueling of AGN}
\label{section:gc}

A number of works have shown that there is disk of young stars around 
Sag A$^\star$ at the Galactic Center with total stellar mass 
$M_\star\sim2\times10^4$\,M$_\odot$,
radius $r\sim0.1$\,pc, scale height of $h\sim0.1r$,
and an average age of $\sim6$\,Myr (Levin \& Beloborodov 2003;
Genzel et al.~2003; Paumard et al.~2006; Lu et al.~2009).
Although the physical origin of these stars is still
unclear, it is possible that the stars formed {\it in situ} 
in a gaseous disk (Levin \& Beloborodov 2003).  
This star formation episode may have 
been accompanied by accretion onto the central black hole, 
perhaps producing an AGN epoch in the Galaxy (e.g., TQM; Nayakshin \& Cuadra 2005;
Levin 2007).  Importantly, its structure may have been determined
by the radiation pressure of starlight on dust grains.

A simple estimate for the minimum star formation 
rate needed to produce the stars that we see now is 
$\dot{M}_\star^{\rm min}\sim{\rm Mass}/{\rm Age}
\sim3\times10^{-3}\,\,{\rm M_\odot \,\,yr^{-1}}$,
corresponding to a star formation rate surface density
of $\dot{\Sigma}_\star^{\rm min}\sim{\dot{M}_\star^{\rm min}}/{\pi r^2}
\sim 10^5\,\,{\rm M_\odot\,\,yr^{-1}\,\,kpc^{-2}}$,
roughly 100 times the surface density of star formation
for the central 100\,pc of Arp 220's western nucleus.
Note that the ratio $L_\star/M_\star$ for a few-Myr old stellar population 
is  $\xi_{1500}\sim1500$ L$_\odot/$M$_\odot$, which 
allows for an order-of-magnitude estimate for the luminosity 
and flux of the stars at formation: 
$L_\star\sim3\times10^7\xi_{1500}$\,L$_\odot$ and $F_\star\sim10^{15}\xi_{1500}$
\,L$_\odot$ kpc$^{-2}$, respectively.

It is interesting to consider the physical conditions 
of the disk {\it in formation} using the observed 
stellar mass density.  We can write down an approximate
lower limit to the total gas mass required to build the stars 
we see today by equating the gas mass and the 
stellar mass: $M_g\sim M_\star$.  This implies a minimum
volumetric and surface gas density of 
$n\sim M_g/(2\pi r^2 h)\sim10^9$\,cm$^{-3}$ and 
$\Sigma_g\sim2\rho h\sim120$\,g cm$^{-2}$ 
$\sim6\times10^5$\,M$_\odot$ pc$^{-2}$,
respectively, assuming that the stars formed
with $h\sim0.1r$.  Note that this estimate for the gas
density of the disk during formation implies an average
dynamical timescale of order $t_{\rm dyn}\sim(G\rho)^{-1/2}\sim3000$\,yr,
which implies that the {\it maximum star formation rate}
exceeds $\dot{M}_\star^{\rm min}$ by a factor of $\sim2000$:
$\dot{M}_\star^{\rm max}\sim6$\,M$_\odot$\,yr$^{-1}$. 

Given these approximate parameters for the disk in
formation, we can ask if it is physically possible for 
a given mechanism to support the gas in vertical 
hydrostatic equilibrium with scale height of order $h$.  
The effective surface temperature of the disk is 
$T_{\rm eff}\sim\left({F}/{\sigma_{\rm SB}}\right)^{1/4}\sim260\,\,{\rm K}$
and the total average optical depth in the vertical direction is 
$\tau\sim600  \kappa_{10} n_9 h_{0.1r}$
where $\kappa_{10}=\kappa_R(T\gtrsim200\,{\rm K})/10$\,cm$^2$ g$^{-1}$.
Note that this implies a midplane temperature of $\sim1300$\,K,
close to --- but somewhat below --- the sublimation temperature 
of dust.  Given $T$ and $n$, the ratio of the 
radiation pressure to the gas pressure is of order $p_r/p_g\sim20$.

\begin{figure}[t]
\centerline{\includegraphics[width=8cm]{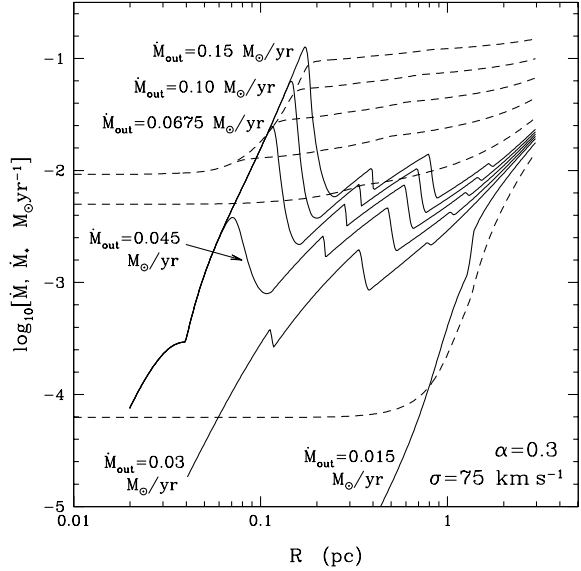}}
\caption{Predicted star formation rate ($\dot{M}_\star$; solid lines) 
and gas accretion rate ($\dot{M}$; dashed lines) as a function of radius 
for conditions appropriate to the Galactic Center, using the model
of Eddington-limited star formation in TQM and discussed in 
\S\ref{section:radiation}  The different line sets indicate different
input gas accretion rate at a fiducial outer radius of 3\,pc,
varying from $\dot{M}_{\rm out}=0.015$\,M$_\odot$ yr$^{-1}$ to 
$0.015$\,M$_\odot$ yr$^{-1}$.  Solutions with higher 
$\dot{M}_{\rm out}$ produce a peak in 
$\dot{\Sigma}_\star$ of $\sim10^5-10^6$\,M$_\odot$
yr$^{-1}$ kpc$^{-2}$ on 0.1\,pc scales.}
\label{gc}
\end{figure}

Hydrostatic equilibrium in the vertical direction requires that 
the radiated flux equal the Eddington flux.  Because the midplane
temperature exceeds $\sim200$\,K, the estimate of equation (\ref{high})
is appropriate and the ratio of the Eddington flux to the 
flux from starlight during formation is 
\begin{equation}
\frac{F_{\rm Edd}}{F_{\star}}\sim
\frac{2\pi G c}{\kappa_R\,\xi}\sim1\,\,\kappa_{10}^{-1}\xi_{1500}^{-1},
\label{ratio_gc}
\end{equation}
where $F_\star\sim10^{15}\xi_{1500}$\,L$_\odot$ kpc$^{-2}$.
The fact that equation (\ref{ratio_gc}) yields a number close to unity
demonstrates the importance of radiation pressure
in this environment for maintaining the disk in vertical hydrostatic
equilibrium.  Although there is evidence for a top-heavy IMF
(Nayakshin \& Sunyaev 2005;
Nayakshin et al.~2006), there is relatively little uncertainty in $\xi$ in 
the above expression relative to the assumption that the disk of 
observed stars can be treated as a disk of gas with approximately 
the same geometry.  It should be noted again that the uncertainty in 
$\kappa_R$ is dominated by the assumption of Galactic dust-to-gas ratio
and Solar metallicity.  If anything, $\kappa_{10}$ may be an 
underestimate of the true opacity.  Nevertheless, equation (\ref{ratio_gc})
highlights the importance of radiation pressure in extreme star formation
environments and may indicate that the formation of the stars in the 
Galactic center region was in fact Eddington-limited (see TQM \& Levin 2007
for more details).

Equation (\ref{ratio_gc}) also shows that 
radiation pressure supported disks may 
vastly exceed a bolometric flux of $\sim10^{13}$\,L$_\odot$ kpc$^{-2}$
if the midplane temperature of the disk is larger than 200\,K
(eq.~\ref{high}).  In the case of the Galactic Center, the flux needed for radiation 
pressure support is of order $F_\star\sim10^{15}$\,L$_\odot$ kpc$^{-2}$,
and this is in good agreement with the inferred flux from the stars
at the Galactic Center.   This is strong evidence that radiation pressure
may be paramount in the self-gravitating disks that accompany AGN fueling.
Indeed, in a detailed study of star formation in local Seyfert nuclei, 
Davies et al.~(2007) find evidence that star formation 
reaches the characteristic flux of equation (\ref{flux})
on $\sim1-10$\,pc scales.   More evidence (besides the Galactic
Center) that the flux from young stars can exceed
$\sim10^{13}$\,L$_\odot$ kpc$^{-2}$ on very small scales
would be useful in constraining detailed models of AGN fueling
(e.g., Levin 2007; Nayakshin et al.~2007).

As an example of a suite of such models, 
Figure \ref{gc} shows the average star formation rate as 
a function of radius from TQM (solid lines) for parameters 
appropriate to the Galactic Center, with a central black hole 
mass of $\sim4\times10^6$\,M$_\odot$
and a bulge velocity dispersion of $\sim75$\,km s$^{-1}$.
For a given input accretion rate at 3\,pc ($\dot{M}_{\rm out}$),
the disk forms stars at the rate required by the Eddington limit.
The dashed lines show the gas accretion rate, which declines
as a function of radius towards the black hole as a result of 
star formation.  Models that fuel the black hole with 
$\sim10^{-2}$\,M$_\odot$ yr$^{-1}$ reach $\dot{\Sigma}_\star
\sim 10^5-10^6$\,M$_\odot$ yr$^{-1}$ kpc$^{-2}$ on 0.1\,pc scales,
producing a sharp peak in the resulting stellar surface density after 
accretion abates (see Appendix A in TQM).\\

{\noindent \bf Other Sources of Feedback:}
In addition to radiation pressure, it is also worth considering other sources of 
pressure support thought to be important in star-forming environments on 
galactic scales: gas, cosmic rays, and magnetic fields (e.g., Boulares \& Cox 1990).  
The gas temperature required to support the disk is 
$T_g\sim{\pi G\Sigma_g^2}/{n k_B} \sim 2\times10^4\,\,{\rm K}$.  
However, if the medium is optically-thick, then $p_r/p_g\sim10^4 T_4^3n^{-1}_9$,
where $T_4=T/10^4$\,K.  Additionally, cooling timescale arguments
and the total energy required from the stellar or AGN 
radiation field show that $T_g$ is difficult to maintain 
(although, see the discussion of $p_g$ in Appendix A of TQM).
The magnetic field strength
required for hydrostatic equilibrium is given by 
$B^2/8\pi\sim\pi G\Sigma_g^2$, which implies $B\sim0.3$\,G 
for fiducial parameters. 
 For cosmic rays, I assume that a
fraction $f$ of the {\it total bolometric output} from the stars
goes into primary cosmic ray protons: $\dot{E}_{\rm CR}\sim f L_\star$,
as perhaps might be provided by stellar winds.
The cooling timescale is dominated by inelastic proton-proton scattering,
with a typical cooling timescale for few-GeV protons of
$t_\pi\sim0.05\,n_9^{-1}$\,yr.  This gives a maximum upper limit to the 
equilibrium midplane CR pressure of $P_{\rm CR}\sim fL_\star t_\pi/(3V)\sim
3\times10^{-6}f$\,ergs cm$^{-3}$, which is very small with  
respect to the total pressure required for hydrostatic equilibrium, 
$\sim3\times10^{-3}$\,ergs cm$^{-3}$.   I conclude that gas pressure
and cosmic ray pressure are both likely sub-dominant with respect to 
radiation pressure in this context, and that magnetic fields of strength 
approaching $\sim$\,G are required to contribute significantly to 
the pressure budget.  

Supernovae are similarly unlikely to 
have supported the disk, given the youth of the observed stellar
population.

\section{Summary \& Conclusions}
\label{section:summary}

Radiation pressure associated with the absorption and scattering of 
starlight by dust grains in rapidly star-forming environments 
is likely to be an important feedback process, and may be responsible for 
the regulation of star formation in some circumstances on galactic 
scales.  Here, I have reiterated and clarified arguments in the 
starburst and AGN contexts developed in TQM.  However, much more
work needs to be done to assess this mechanism in full.  In particular,
it is crucial to understand the coupling between the radiation field 
produced by spatially clustered massive stars and the highly 
inhomogeneous and turbulent ISM of starbursts (e.g., 
Krumholz \& Matzner 2009; Murray et al.~2009).  This remains
a primary avenue of future investigation.\\

\noindent{\bf Acknowledgments:}  I thank E.~Quataert and N.~Murray
for numerous stimulating discussions and for collaboration on the
projects and ideas described here.  Support
for this effort is provided in part by an Alfred P.~Sloan Fellowship.

\end{document}